\documentclass[useAMS]{mn2e}
\usepackage{graphicx}
\usepackage[fleqn]{amsmath}

\title{The dynamics of eccentric accretion discs in superhump systems}
\author[S.G. Goodchild, G.I. Ogilvie]{Simon Goodchild$^{1}$\thanks{E-mail: sgg@ast.cam.ac.uk (SGG)} and Gordon Ogilvie$^{1,2}$\\
$^{1}$Institute of Astronomy, Madingley Road, Cambridge CB3 0HA\\$^{2}$ Department of Applied Mathematics and Theoretical Physics, Wilberforce Road, Cambridge CB3 0WA}

\begin{document}
\maketitle

\newcommand{\rme}{\mathrm{e}}
\newcommand{\rmi}{\mathrm{i}}
\newcommand{\rmb}{\mathrm{b}}
\newcommand{\rmd}{\mathrm{d}}

\begin{abstract}
We have applied an eccentric accretion disc theory in simplified form to the case of an accretion disc in a binary system, where the disc contains the 3:1 Lindblad resonance. This is relevant to the case of superhumps in SU Ursae Majoris cataclysmic variables and other systems, where it is thought that this resonance leads to growth of eccentricity and a modulation in the light curve due to the interaction of a precessing eccentric disc with tidal stresses. A single differential equation is formulated which describes the propagation, resonant excitation and viscous damping of eccentricity. The theory is first worked out in the simple case of a narrow ring and leads to the conclusion that the eccentricity distribution is locally suppressed by the presence of the resonance, creating a dip in the eccentricity at the resonant radius. Application of this theory to the superhump case confirms this conclusion and produces a more accurate expression for the precession rate of the disc than has been previously accomplished with simple dynamical estimates.
\end{abstract}

\section{Introduction}
\label{intr}

A superhump is a periodic variation in the luminosity of an accreting binary system, with a period close to the orbital period of the binary. Superhumps were first discovered in the SU Ursae Majoris class of dwarf novae, where they occur during superoutbursts which are longer and more pronounced than common outbursts. These usually show so-called 'normal' or positive superhumps, where the superhump period is slightly longer than the orbital period. Negative superhumps, where the superhump period is shorter than the orbital period, have also been observed. Some other binary systems, such as V603 Aquilae (Patterson \& Richman 1991), have been described as permanent superhump systems, where the superhumps do not occur in outbursts but are instead variations in the normal lightcurve of the systems. Superhumps have also been observed in low-mass X-ray binaries (LMXBs), for example KV UMa/XTE J1118+480 (Uemura et al 2002).

The most probable explanation for superhumps in the SU UMa systems is in terms of the effect of tidal stresses on a precessing, eccentric accretion disc, a model first proposed for the SU UMa systems by Whitehurst (1988) and investigated by Hirose \& Osaki (1990) and Lubow (1991). This last paper derived an analytic expression for an instability at the 3:1 Lindblad resonance that was first noted by Whitehurst, which leads to eccentricity growth in a previously circular disc. Haswell et al. (2001) proposed a similar explanation for the superhumps in LMXBs, in which the precessing disc leads to a modulation in the light from reprocessed X-rays from the central source.

Observational support for the idea of an eccentric disc first came from eclipse maps of dwarf novae (Hessman et al. 1992; Rolfe, Haswell \& Patterson 2000) and has recently been added to by the spectroscopic study of Zhao et al. (2005). Many numerical models of eccentric discs in cataclysmic variables have also been studied. Smoothed-particle hydrodynamics codes (e.g. Hirose \& Osaki 1990; Lubow 1992; Murray 1998) in general produce eccentric discs and superhumps, although some Eulerian codes do not (e.g. Heemskerk 1994;  Kornet \& Rozyczka 2000).

Theoretically, the precession rate of an eccentric disc has traditionally been estimated using the dynamical precession frequency $\omega_{\mathrm{dyn}}$ of a particle due to tidal forces. This is given by the expression
\begin{equation}
\label{pra}
\frac{\omega_{\mathrm{dyn}}}{\Omega_{\rmb}}=\frac{q}{(1+q)^{1/2}} \left( \frac{1}{4} \left(\frac{r}{a}\right)^{1/2} b^{(1)}_{3/2}\left(\frac{r}{a}\right) \right)
\end{equation}
as a fraction of the orbital frequency of the binary $\Omega_{\rmb}$, where $a$ is the semi-major axis of the binary and $b^{(1)}_{3/2}$ is a Laplace coefficient (see equation (\ref{lap}) below). Observationally, the precession rate is usually expressed in terms of the superhump period excess $\varepsilon$, defined as
\begin{equation}
\label{exc}
\varepsilon=\frac{\omega_{\mathrm{prec}}}{\Omega_{\rmb}-\omega_{\mathrm{prec}}}.
\end{equation}
Mass ratio and precession rate should thus be roughly proportional, since equation (\ref{exc}) can be approximated as
\begin{equation}
\varepsilon \simeq \frac{\omega_{\mathrm{prec}}}{\Omega_{\rmb}}+\left(\frac{\omega_{\mathrm{prec}}}{\Omega_{\rmb}}\right)^{2}
\end{equation}
for the low values of $q$ found in SU UMa stars (Patterson et al (2005) found $q<0.35$).

Observations indeed show an approximately linear relation between $q$ and $\varepsilon$, and for example Patterson (2001) found
\begin{equation}
\varepsilon=(0.216\pm0.018)q.
\end{equation}
As pointed out by Murray (2000), although equation (\ref{pra}) reproduces the correct behaviour of disc precession period with mass ratio, it does not accurately reproduce the constant of proportionality, which is about 0.4 if the dynamical precession is evaluated at the 3:1 Lindblad resonance itself. To obtain a more accurate expression, we therefore need a fluid dynamical treatment of the disc. The overestimation may be due to a subtractive term in the relation due to pressure effects, as argued by Murray (2000) using the analysis of Lubow (1992). Alternatively, Patterson (2001) noted that the correct proportionality can be reproduced simply by changing the radius at which equation (\ref{pra}) is evaluated (with $\langle r\rangle\simeq0.37a$ to match the observations) and suggested that the precession rate may be determined by the effect of the tidal potential on a larger disc annulus rather than just the resonance. The actual precession would then be an average in some way over the disc.

The aim of this paper is to construct a more complete description of the behaviour of eccentricity in the accretion discs of SU UMa systems, using the theory of eccentric discs derived by Ogilvie (2001). The following section details the derivation of an equation for the eccentricity in a 2D disc, including the effects of the tidal precession due to the binary companion, eccentricity excitation at the 3:1 Lindblad resonance and damping of eccentricity due to viscosity. Section \ref{soln} then describes solutions of this equation in the form of normal modes for both a narrow ring and a full disc, which give values for the disc precession rate and the growth rate of eccentricity. These results are summarised and discussed in section \ref{diss}.

\section{The eccentricity equation for a 2D disc}
\label{derv}

\subsection{Derivation of the eccentricity equation}

To derive an equation for the eccentricity in a disc, we first consider the unperturbed disc, using a simple two dimensional, inviscid model. In terms of the polar velocity components (\textit{u},\textit{v}), pressure $p$, density $\rho$, adiabatic exponent $\gamma$ and gravitational potential $\Phi$, the equations of motion are
\begin{gather}
\label{ram}
\frac{\partial u}{\partial t}+u\frac{\partial u}{\partial r}+\frac{v}{r}\frac{\partial u}{\partial \phi}-\frac{v^{2}}{r} =-\frac{1}{\rho}\frac{\partial p}{\partial r}-\frac{\partial \Phi}{\partial r} \\
\label{anm}
\frac{\partial v}{\partial t}+u\frac{\partial v}{\partial r}+\frac{v}{r}\frac{\partial v}{\partial \phi}+\frac{uv}{r} =-\frac{1}{r\rho}\frac{\partial p}{\partial \phi}-\frac{1}{r}\frac{\partial \Phi}{\partial \phi} \\
\label{mac}
\frac{\partial \rho}{\partial t}+u\frac{\partial \rho}{\partial r}+\frac{v}{r}\frac{\partial \rho}{\partial \phi} =-\frac{\rho}{r} \left(\frac{\partial (ru)}{\partial r}+\frac{\partial v}{\partial \phi} \right)\\
\label{enc}
\frac{\partial p}{\partial t}+u\frac{\partial p}{\partial r}+\frac{v}{r}\frac{\partial p}{\partial \phi} =-\frac{\gamma p}{r} \left(\frac{\partial (ru)}{\partial r}+\frac{\partial v}{\partial \phi} \right).
\end{gather}

These equations are solved by treating each quantity as a function of a parameter $\epsilon$, which is a characteristic value of the disc semi-thickness $H/r$ and is assumed to be small. Quantities can be expanded in terms of $\epsilon$ as power series
\begin{equation}
x=x_{0}+x_{2}\epsilon^{2}+\cdots
\end{equation}
and the terms at each order in the equations equated.

In the basic state of the disc, we assume that the disc is steady, axisymmetric, non-self-gravitating and has no radial motion. We also assume vertical hydrostatic equilibrium (valid in some sense if the 2D quantities are treated as 3D quantities integrated vertically through the disc), which leads to a scaling $p(r)\sim\epsilon^{2}\rho(r)$ between the pressure and density and therefore a factor of $\epsilon^{2}$ in the expansion for the pressure. Simplifying equations (\ref{ram})-(\ref{enc}) for this basic state, we obtain
\begin{gather}
r\Omega_{0}^{2}=\frac{\partial\Phi_{0}}{\partial r}\\
\label{omt}
2r\Omega_{0} \Omega_{2}=\frac{\partial\Phi_{2}}{\partial r}+\frac{1}{\rho_{0}}\frac{\partial p_{0}}{\partial r}
\end{gather}
for the angular velocity $\Omega$, where $\Phi_{0}$ is the potential for a point mass and $\Phi_{2}$ is an axisymmetric correction due to the binary companion, taken to be of $O(\epsilon^{2})$ since it is small compared to the primary potential.

We then perturb the fluid equations, introducing a linear, adiabatic perturbation $x'(r)\rme^{-\rmi\phi}$ for each quantity. The perturbations satisfy the equations
\begin{gather}
\label{rap}
\frac{\partial u'}{\partial t}-\rmi\Omega u'-2\Omega v' =-\frac{1}{\rho}\frac{\partial p'}{\partial r}+\frac{\rho'}{\rho^{2}}\frac{\partial p}{\partial r} \\
\label{anp}
\frac{\partial v'}{\partial t}-\rmi\Omega v'+\frac{u'}{r}\frac{\partial}{\partial r}\left(r^{2}\Omega\right)=\frac{\rmi p'}{r\rho} \\
\label{map}
\frac{\partial \rho'}{\partial t}-\rmi\Omega \rho'+u'\frac{\partial \rho}{\partial r}=-\frac{\rho}{r} \left(\frac{\partial (ru')}{\partial r}-\rmi v' \right)\\
\label{enp}
\frac{\partial p'}{\partial t}-\rmi\Omega p'+u'\frac{\partial p}{\partial r} =-\frac{\gamma p}{r} \left(\frac{\partial (ru')}{\partial r}-\rmi v' \right).
\end{gather}
We expand the perturbed quantities in the same way as the basic state:
\begin{equation}
x'=x'_{0}+x'_{2}\epsilon^{2}+\cdots
\end{equation}
Since we are interested the evolution of the disc on time-scales slower than the orbital time-scale, we also introduce a slow time variable $\tau$, related to the time coordinate as $\tau=\epsilon^{2} t$. We do not need to expand the potential, since, for a circular binary system over period of time, the average potential is axisymmetric.

At lowest order, equations (\ref{rap}) and (\ref{anp}) yield the relations
\begin{gather}
-\rmi\Omega_{0} u'_{0}-2\Omega_{0}v'_{0}=0 \\
-\rmi\Omega_{0} v'_{0}+\frac{1}{2}\Omega_{0}u'_{0}=0.
\end{gather}
Therefore $2v'_{0}=-\rmi u'_{0}$, and we can set
\begin{gather}
u'_{0}=\rmi r\Omega_{0}E(r,\tau) \\
v'_{0}=\frac{1}{2}r\Omega_{0}E(r,\tau).
\end{gather}
By comparison with the expressions for velocities in an elliptical Keplerian orbit, we see that, for small eccentricities, this describes a Keplerian ellipse, and the function $E(r,\tau)$ is the complex eccentricity $E=e \rme^{\rmi\varpi}$ of the orbit. This solution arises because pressure and tidal effects do not enter the equation at this order, and we obtain the expected solution for a free particle in a Keplerian potential.

By taking an appropriate linear combination of equations (\ref{rap}) and (\ref{anp}) at $O(\epsilon^{2})$, $u'_{2}$ and $v'_{2}$ can be eliminated and an equation for the eccentricity derived. We use equations (\ref{map}) and (\ref{enp}) at lowest order to eliminate $\rho'_{0}$ and $p'_{0}$, and equation (\ref{omt}) to eliminate $\Omega_{2}$ in favour of the tidal potential. Putting this together and removing the scaling factors, we obtain a partial differential equation for the complex eccentricity
\begin{equation} \label{ecc}
2r\Omega\frac{\partial E}{\partial t}=\frac{-\rmi E}{r} \frac{\partial}{\partial r} \left(r^{2} \frac{\partial \Phi_{2}}{\partial r} \right)+\frac{\rmi E}{\rho} \frac{\partial p}{\partial r}+\frac{\rmi}{r^{2}\rho}\frac{\partial}{\partial r} \left( \gamma p r^{3} \frac{\partial E}{\partial r} \right).
 \end{equation}

The form of this equation without the tidal potential already sheds some light on the nature of the various solutions. We can Fourier analyse in time by setting $E(r,t)=E(r)\rme^{\rmi\omega t}$. The eigenvalue $\omega$ is determined by the Sturm-Liouville equation
\begin{equation} \label{ecs} \begin{split}
2r\Omega\omega E = \frac{E}{\rho} \frac{\partial p}{\partial r}+\frac{1}{r^{2}\rho}\frac{\partial}{\partial r} \left( \gamma p r^{3} \frac{\partial E}{\partial r} \right).
\end{split} \end{equation}
 Solutions of this equation with appropriate boundary conditions lead to a series of normal modes in the disc. Each mode is a radial eccentricity distribution which precesses at a frequency given by its eigenvalue.

\subsection{The tidal potential}
\label{perp}

To make further progress, we require the tidal potential due to a binary companion. As we are interested in long-term evolution, we can use the Gauss averaging method (e.g Murray \& Dermott 1999) where the companion is treated as a smeared-out ring of material along its orbit, with the density proportional to the time taken to traverse the element. For SU UMa systems the orbit of the companion is circular, and thus the potential is given by the expression
\begin{equation}
\label{per}
\Phi(r,\phi)=\int_{0}^{2\pi}\frac{-GM_{2} \mathrm{d}\phi_{\mathrm{b}}}{2\pi (a^{2}+r^{2}-2ra\cos(\phi-\phi_{\rmb}))^{1/2}}
\end{equation}
for a companion of mass $M_{2}$ at polar coordinates $(a,\phi_{\rmb})$. This integral can be written in a shortened form, using a Laplace coefficient from celestial mechanics, as
\begin{equation} \label{peb}
\Phi(r,\phi) = -\frac{GM_{2}}{2a}b^{(0)}_{1/2}\left(\frac{r}{a}\right),
\end{equation}
where the Laplace coefficient $b^{(j)}_{s/2}$ is defined as
\begin{equation}
\label{lap}
b^{(j)}_{s/2}(\alpha)=\frac{1}{\pi}\int_{0}^{2 \pi} \frac{\cos (j\psi) \mathrm{d}\psi}{(1+\alpha^{2}-2\alpha\cos\psi)^{s/2}}.
\end{equation}
This is then inserted into equation (\ref{ecc}). After some manipulation and simplification using identities for the derivatives of Laplace coefficients, it becomes
\begin{equation} \label{ecb} \begin{split}
2r\Omega\frac{\partial E}{\partial t} &=\frac{\rmi E}{\rho} \frac{\partial p}{\partial r}+\frac{\rmi}{r^{2}\rho}\frac{\partial}{\partial r} \left( \gamma p r^{3} \frac{\partial E}{\partial r} \right)+\\& \frac{\rmi q \Omega^{2} r^{3}}{2a^{2}}\left( b^{(1)}_{3/2}\left(\frac{r}{a}\right)E\right),
\end{split} \end{equation}
where $q=M_{2}/M_{1}$ is the mass ratio of the binary. The new term is also proportional to \textit{E} and simply modifies the Sturm-Liouville eigenvalue problem of the previous section, leading to a new eigenvalue and corresponding eigenfunction.

\subsection{Generation and damping of eccentricity}
\label{geda}

The Sturm-Liouville problem of section \ref{derv} describes how an eccentric disc evolves, but does not consider how the eccentricity in the disc is produced or describe whether it can be damped. Eccentricity in the discs of SU UMa systems is thought to be due to an instability at the 3:1 Lindblad resonance, and an analytical expression for the eccentricity growth rate $\xi$ was found by Lubow (1991):
\begin{equation}
\xi=2.08C\Omega_{\rmb}q^{2}\frac{r_{\mathrm{res}}}{r_{o}-r_{i}}
\end{equation}
where the binary has an angular velocity $\Omega_{\rmb}$ and \textit{C} is a correction factor dependent on the size of the ring. We insert this growth term into the equation describing the eccentricity as a delta-function at the resonance point, so the expression describes the rate at which eccentricity is created by the resonance and the fluid dynamics of the eccentricity equation itself describe how it propagates through the disc. The correction factor \textit{C}, which describes the effect of finite disc width on the growth rate, is therefore 1 and the denominator is replaced by a delta-function in the numerator. Thus the modified form of equation (\ref{ecb}) is
\begin{equation} \label{ecr} \begin{split}
2r\Omega\frac{\partial E}{\partial t} &=\frac{\rmi E}{\rho} \frac{\partial p}{\partial r}+\frac{\rmi}{r^{2}\rho}\frac{\partial}{\partial r} \left( \gamma p r^{3} \frac{\partial E}{\partial r} \right)+\\& \frac{\rmi q \Omega^{2} r^{3}}{2a^{2}}\left( b^{(1)}_{3/2}\left(\frac{r}{a}\right)E\right)+2\xi r\Omega E \delta(r-r_{\mathrm{res}}).
\end{split} \end{equation}

When analysed as an eigenvalue problem, the eigenvalue is now complex, with a real part describing the precession frequency of the disc, and an imaginary part which determines whether the mode grows or decays in time. To analyse the eccentricity growth more completely, we introduce a viscosity to the disc, which would be expected to compete with the resonant growth and cause damping of the eccentricity. Ogilvie (2001) demonstrated that a Navier-Stokes shear viscosity alone results in an overstability of the 2D disc, and since we wish to consider damping of eccentricity, we work with a bulk viscosity rather than a shear term. Thus we introduce a bulk viscosity described by a Shakura-Sunyaev (1973) $\alpha$-parameter. The equations of motion of the disc (\ref{ram})--(\ref{anm}) become
\begin{gather}
\label{ras}
\frac{\partial u}{\partial t}+u\frac{\partial u}{\partial r}+\frac{v}{r}\frac{\partial u}{\partial \phi}-\frac{v^{2}}{r} =-\frac{1}{\rho}\frac{\partial p}{\partial r}-\frac{\partial \Phi}{\partial r}+\frac{1}{\rho}\frac{\partial T}{\partial r} \\
\label{ans}
\frac{\partial v}{\partial t}+u\frac{\partial v}{\partial r}+\frac{v}{r}\frac{\partial v}{\partial \phi}+\frac{uv}{r} =-\frac{1}{r\rho}\frac{\partial p}{\partial \phi}-\frac{1}{r}\frac{\partial \Phi}{\partial \phi}+\frac{1}{r\rho}\frac{\partial T}{\partial \phi}
\end{gather}
where the quantity \textit{T} describing the bulk viscosity is given by
\begin{equation} \label{bul}
T=\frac{\alpha_{\rmb} p}{r\Omega}\left(\frac{\partial (ru)}{\partial r}+\frac{\partial v}{\partial \phi} \right).
\end{equation}
Since this viscous term is proportional to the pressure, we scale it to the same order $O(\epsilon^{2})$ and the lowest order equations are unaffected. Therefore the perturbed flow can still be described as an eccentricity, now governed by a new eccentricity equation. Repeating the analysis that led to equation (\ref{ecc}), we find that the bulk viscosity has the simple effect of changing the adiabatic index $\gamma$ to a complex number;
\begin{equation}
\gamma \to \gamma-\rmi\alpha_{\rmb}.
\end{equation}
Therefore the full equation describing generation, damping and dynamics of eccentricity in the disc is
\begin{equation} \label{ecf} \begin{split}
2r\Omega\frac{\partial E}{\partial t} &=\frac{\rmi E}{\rho} \frac{\partial p}{\partial r}+\frac{\rmi}{r^{2}\rho}\frac{\partial}{\partial r} \left((\gamma-\rmi\alpha_{\rmb}) p r^{3} \frac{\partial E}{\partial r} \right)+\\& \frac{\rmi q \Omega^{2} r^{3}}{2a^{2}}\left( b^{(1)}_{3/2}\left(\frac{r}{a}\right)E\right)+2\xi r\Omega E \delta(r-r_{\mathrm{res}}).
\end{split} \end{equation}
Splitting this into its constituent terms, we can interpret the effect of each on the dynamics of eccentricity. The term proportional to $\gamma$ has the form of a dispersive wave equation like the Schr\"odinger equation, allowing the eccentricity to propagate through the disc as a dispersive wave. By analogy with the integrated probability density in the Schr\"odinger equation, this dispersion has a conserved quantity 
\begin{equation}
\label{con}
\frac{\partial}{\partial t}\int_{r_{\mathrm{in}}}^{r_{\mathrm{out}}}\frac{1}{2}r^{3}\rho\Omega |E|^2 \mathrm{d}r=0
\end{equation}
with the boundary condition assumed below ($p \partial E/\partial r=0$ at the boundaries). The dynamical term simply describes a precession of the eccentricity which also satisfies equation (\ref{con}). The last term describes the eccentricity growth, and the term proportional to the bulk viscosity describes diffusion of eccentricity through the disc. These terms now break the conservation relation of equation (\ref{con}), and we have the more general relation
\begin{gather}
\frac{\partial}{\partial t}\int_{r_{\mathrm{in}}}^{r_{\mathrm{out}}}\frac{1}{2}r^{3}\rho\Omega |E|^2 \mathrm{d}r=\nonumber \\ \label{cos} \xi\rho r^3\Omega|E|^2\Bigg|_{\mathrm{res}}-\int_{r_{\mathrm{in}}}^{r_{\mathrm{out}}}\frac{1}{2}\alpha_{\rmb}p r^3\left|\frac{\partial E}{\partial r}\right|^2{\rm d}r.
\end{gather}
The quantity on the left hand side of equations (\ref{con})--(\ref{cos}) is the angular momentum deficit of the disc, which expresses the difference between the angular momentum of elliptical and circular orbits with the same energy.

As before, we can Fourier analyse in time to obtain an eigenvalue problem similar to a Sturm-Liouville problem, but with complex eigenvalues. This still has a series of modes, which can be characterised by their number of `nodes' or points where the phase of the eccentricity passes sharply through $\pi$.
A useful expression for this eigenvalue can be derived  by multiplying equation (\ref{ecf}) by $r^{2}\rho E^{*}$ and integrating over the radial extent of the disc. The term in the eccentricity gradient can be integrated by parts and, with the boundary conditions we use below, the boundary term vanishes and we obtain expressions containing integrals of real quantities only. Therefore we find the following expressions for the real and imaginary parts of the eigenvalue:
\begin{gather}
\label{amd}
\Re(\omega) = \frac{\int r^{2}\frac{\partial p}{\partial r} |E|^{2} \rmd r-\int\gamma r^{3}p \big\vert\frac{\partial E}{\partial r}\big\vert^{2} \rmd r+\int\frac{\rho q r^{5} \Omega^{2}}{2a^{2}}b_{3/2}^{(1)} |E|^{2} \rmd r}{2\int r^{3}\rho\Omega |E|^{2} \mathrm{d}r} \\
\label{amr}
\Im(\omega) = \frac{\int\alpha_{\rmb} r^{3}p \big\vert\frac{\partial E}{\partial r}\big\vert^{2} \rmd r-2\xi\Omega r^{3} \rho|E|^{2}|_{\mathrm{res}}}{2\int r^{3}\rho\Omega |E|^{2} \mathrm{d}r}.
\end{gather}

These expressions show that there is a separation of effects into those which affect the precession rate, described by the real part of the eigenvalue, and those which affect the growth rate, which is determined by the imaginary part. The real part has three separate contributions. The first two are due to pressure, and if the pressure gradient is negative (as in the disc models we used) both of these terms are negative and lead to a retrograde precession. The potential term is positive and induces a prograde precession, so these three terms will compete to determine the direction of overall disc precession. 
The two terms in the imaginary part behave as expected, with the resonant forcing acting to increase eccentricity and the bulk viscosity damping it (a negative eigenvalue implies a growing eccentricity). Equation (\ref{amr}) is simply another form of the general relation for eccentricity growth and decay expressed by equation (\ref{cos}), in terms of a growth-rate eigenvalue.

\section{The solutions of the eccentricity equations}
\label{soln}

\subsection{Behaviour of eccentricity in a narrow ring}
\label{nari}

The overall equation for the eccentricity can be simplified by considering a narrow ring centred on the 3:1 Lindblad resonance. In this case, the pressure and density can be assumed to be approximately constant and the tidal term due to the companion can be ignored as it causes a constant precession, leaving only the eccentricity gradient and the resonant term. Equation (\ref{ecf}) therefore greatly simplifies to the form
\begin{equation} \label{nare}
\frac{\partial E}{\partial t}=\frac{\rmi(\gamma-\rmi\alpha_{\rmb})p}{2 \rho \Omega}\frac{\partial^{2} E}{\partial x^{2}}+\xi E\delta(x)
\end{equation}
where $x=r-r_{\mathrm{res}}$.
This is helpful because the equation now has an analytical solution and the behaviour of the eccentricity can be examined directly. The parameter $\xi$ can be varied to study the effect of changing the resonant strength.
Equation (\ref{nare}) can be solved as an eigenvalue problem with an eigenfrequency $\omega$ as before. We solve the equations in a ring extending between $x=-a$ and $x=b$, with the boundary conditions that the eccentricity gradient is zero at the edges of the ring. The actual numerical value of the eccentricity is multiplied by an arbitrary constant, and we scale by setting eccentricity at the inner edge $-a$ to 1; since an absolute eccentricity$>$1 describes an open orbit, eccentricities greater than 1 imply that the eccentricity at the inner edge must be low. The solutions are
\begin{gather}
E(x)=\cos k(x+a), \quad -a \leq r<0\nonumber \\
E(x)=\frac{\cos ka}{\cos kb}\cos k(x-b), \quad 0<r \leq b,
\end{gather}
The wavenumber \textit{k} is complex and satisfies the equation
\begin{equation} \label{keq}
k(\tan(ka)+\tan(kb))=\frac{2\rmi \xi\rho\Omega}{p(\gamma-\rmi\alpha_{\rmb})}=\frac{\rmi W}{(a+b)(\gamma-\rmi\alpha_{\rmb})}
\end{equation}
where we have introduced the dimensionless parameter $W$ to characterise the strength of the resonance. The eigenfrequency is then given by
\begin{equation} \label{disp}
\omega=-\frac{(\gamma-\rmi \alpha_{\rmb})p}{2\rho\Omega}k^{2}.
\end{equation}
With no resonance, the eccentricity distributions are simple cosine waves. When the resonance is added a cusp develops at the resonance point in the centre of the ring, with a discontinuity in the eccentricity gradient. The behaviour of this cusp as the resonant strength increases is dependent on the behaviour of equation (\ref{keq}) which can be expanded and solved analytically for small and large values of $W$. For $W\ll1$, $k$ or $\tan(ka)$ (and therefore both for the lowest order solution) must be small and therefore we can write $\tan(ka)\approx ka$. Therefore the wavenumber is approximately
\begin{equation}
k\simeq \sqrt{\frac{2\rmi \xi\Omega\rho}{p(\gamma-\rmi\alpha_{\rmb})(a+b)}},
\end{equation}
with a phase depending on the ratio of $\gamma$ to $\alpha$. Equation (\ref{disp}) at lowest order in $k$ gives an approximately imaginary eigenvalue
\begin{equation}
\omega\simeq -\frac{\rmi \xi}{a+b},
\end{equation}
in agreement with the growth rates derived for narrow rings by Lubow (1991). The following term in the series gives a more accurate expression for the growth rate
\begin{equation}
\Im(\omega)=-\frac{\xi}{a+b}+\frac{2\xi^{2}\Omega\alpha_{\rmb}\rho(a^{3}+b^{3})}{3p(\gamma^{2}+\alpha_{\rmb}^{2})(a+b)^{3}},
\end{equation}
and an expression for the precession frequency of the disc
\begin{equation}
\Re(\omega)=-\frac{2\xi^{2}\Omega\gamma\rho(a^{3}+b^{3})}{3p(\gamma^{2}+\alpha_{\rmb}^{2})(a+b)^{3}},
\end{equation}
which is purely retrograde and pressure-driven, as the tidal potential of the companion has been ignored.

\begin{figure}
\centering
\includegraphics[angle=0, width=1\columnwidth]{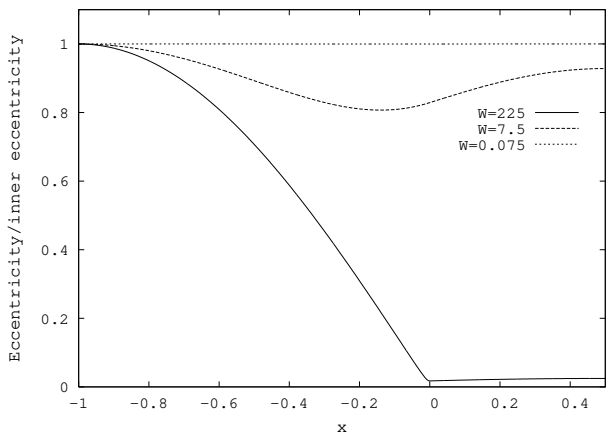}
\caption{The eccentricity of narrow rings with different resonant strengths, showing the development of the cusp as $W$ increases. Ring dimensions $a=1, b=0.5$.}
\label{fig:narw}
\end{figure}

\begin{figure}
\centering
\includegraphics[angle=0, width=1\columnwidth]{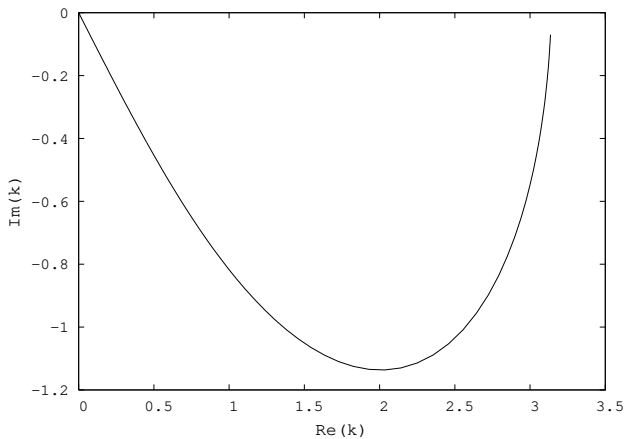}
\caption{Argand diagram for the eccentricity wavenumber with increasing resonant strength. Ring dimensions $a=1, b=0.5$.}
\label{fig:karg}
\end{figure}

\begin{figure}
\centering
\includegraphics[angle=0, width=1\columnwidth]{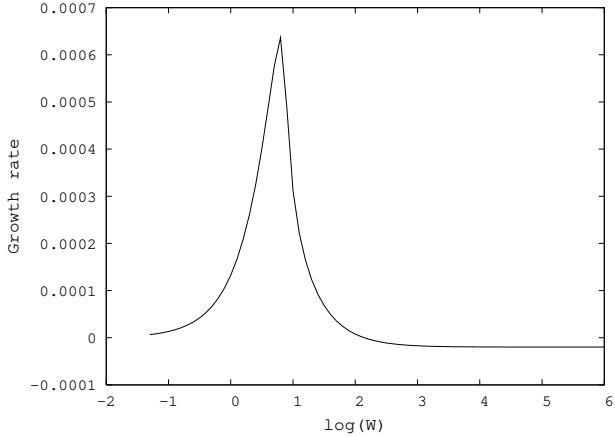}
\caption{The dependence of eccentricity growth on resonance strength for a ring with $a=1, b=0.5, \alpha_{\rmb}=0.1$. Growth rate expressed in terms of the orbital frequency.}
\label{fig:narr}
\end{figure}

For a very strong resonance with $W\gg1$, the dominant terms are $\tan(ka)$ or $\tan(kb)$, which can become arbitrarily large as $ka$ or $kb$ approaches $(n+1/2)\pi$. This leads to a lowest-order solution in which (assuming $a>b$)
\begin{equation}
k\simeq\frac{\pi}{2a}
\end{equation}
with the eccentricity strongly suppressed at the resonant radius, such as shown in Fig. \ref{fig:narw} for $W=225$. As the resonance is made stronger the cusp becomes deeper and the wavenumber closer to the value where the eccentricity has a zero at the resonant point. The growth rate converges on a constant, since the wavenumber is now approaching a single value, and is given by
\begin{equation}
\Im(\omega)\simeq\frac{\alpha_{\rmb} p \pi^{2}}{8\Omega\rho a^{2}}.
\end{equation}
In between these two extremes the wavenumber behaves as shown in Fig. \ref{fig:karg}, which was produced by solving equation (\ref{keq}) numerically, initially increasing before converging on the value of the high-strength limit. The behaviour of the growth rate in terms of $W$ is shown in Fig. \ref{fig:narr} and shows a maximum at a value of about $W=10$, above which the growth rate falls and the mode stabilises at about $10^{2}$. Increasing the growth rate therefore ultimately leads to a stable mode rather than a highly unstable one. This counter-intuitive behaviour arises because of the cusp, which, as shown by Fig. \ref{fig:narw}, deepens as the growth rate increases and leads to a low eccentricity at the resonant point. Since the eccentricity growth here is proportional to the eccentricity itself, a low eccentricity slows the growth of the instability and allows the eccentricity to be damped away in the disc.

\begin{figure}
\centering
\includegraphics[angle=0, width=1\columnwidth]{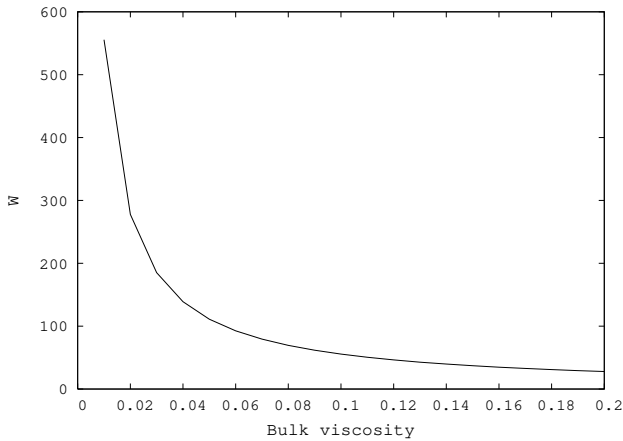}
\caption{The marginal stability curve for a narrow ring with $a=1,$, $b=0.5$ in the plane $W-\alpha_{\rmb}$.}
\label{fig:stab}
\end{figure}

Whether the eccentricity is able to grow is determined by the parameters $\alpha_{\rmb}$ and $W$. Fig. \ref{fig:stab} shows the marginal stability curve in the plane $W$-$\alpha_{\rmb}$, with values below the curve shown being unstable. At high $W$, the disc is unstable only for very low values of the viscosity, whereas it is unstable for most viscosities at lower $W$.

\subsection{Solutions of the full eccentricity equation-precession}
\label{stso}

We solved the eccentricity equations (\ref{ecs}), (\ref{ecb}) and (\ref{ecf}) for their normal modes numerically using a 4th/5th order Runge-Kutta method with adaptive step-size. The pressure $p$ and density $\rho$ of the two-dimensional disc in section \ref{derv} are interpreted as the vertically integrated pressure $P$ and surface density $\Sigma$ of a three-dimensional disc. In solving the equations, the distributions chosen were for a steady alpha-disc with Kramers opacity, as described, for example, in Frank, King \& Raine (2002).  These have the form
\begin{gather}
P=P_{\mathrm{sc}}\left(\frac{r}{r_{\mathrm{sc}}}\right)^{-3/2}\left(1-\sqrt{\frac{r_{\mathrm{in}}}{r}}\right)\tanh\left(\frac{r_{\mathrm{out}}-r}{\nu r_{\mathrm{sc}}}\right)\\
\Sigma=\Sigma_{\mathrm{sc}}\left(\frac{r}{r_{\mathrm{sc}}}\right)^{-3/4}\left(1-\sqrt{\frac{r_{\mathrm{in}}}{r}}\right)^{0.7}\tanh\left(\frac{r_{\mathrm{out}}-r}{\nu r_{\mathrm{sc}}}\right),
\end{gather}
where $P_{\mathrm{sc}}$ and $\Sigma_{\mathrm{sc}}$ are the pressure and surface density at the scaling radius $r_{\mathrm{sc}}$. The $\tanh$ functions implement the boundary conditions that pressure and surface density are zero at the outer edge (they are naturally zero at the inner edge in this model) and contain a parameter $\nu$ to allow the width of the drop to 0 to be varied. In all cases, the adiabatic index $\gamma=5/3$.

We describe the disc properties using a characteristic value of the ratio of the disc semi-thickness to the radius, $H/r$. The disc scale height $H$ satisfies the relation
\begin{equation}
P=\Sigma \Omega^{2}H^{2}.
\end{equation}
The ratio $H/r$ depends weakly on radius, given by the expression
\begin{equation}
\label{heq}
\frac{H}{r}=h \left(1-\sqrt{\frac{r_{\mathrm{in}}}{r}}\right)^{3/20}\left(\frac{r}{r_{\mathrm{in}}} \right)^{1/8},
\end{equation}
where the characteristic value $h$ depends on the system parameters shear viscosity $\alpha$, mass transfer rate $\dot{M}$ and white dwarf mass $M_{1}$ and radius $r_{\mathrm{in}}$ as
\begin{gather}
h=\left(\frac{P_{\mathrm{sc}}}{\Sigma_{\mathrm{sc}}}\right)^{1/2}\frac{1}{r_{\mathrm{sc}}\Omega_{\mathrm{sc}}}=\nonumber\\
0.010 \alpha^{-1/10} \left(\frac{\dot{M}}{10^{16} \mathrm{gs}^{-1}}\right)^{3/20} \left(\frac{M_{1}}{M_{\odot}}\right)^{-3/8} \left(\frac{r_{\mathrm{in}}}{8\times10^{8}\mathrm{cm}}\right)^{1/8}.
\end{gather}
The constant was evaluated using a numerical solution of a disc with full vertical structure as described by Ogilvie and Dubus (2001).

A particular binary system is characterised by its mass ratio $q$ and semi-major axis $a$, while the disc itself is described by the characteristic disc semi-thickness $h$ and the bulk viscosity parameter $\alpha_{\rmb}$. The radius of the 3:1 Lindblad resonance is given by the expression
\begin{equation}
r_{\mathrm{res}}=3^{-2/3}(1+q)^{-1/3}a
\end{equation}
and the outer radius of the disc was evaluated using the prescription of Whitehurst and King (1991), in which the disc is truncated at 0.9 of the Roche radius as given by the Eggleton (1983) approximation, so
\begin{equation}
r_{\mathrm{out}}=\frac{0.45a}{0.6+q^{2/3}\log(1+q^{-1/3})}.
\end{equation}

The first set of solutions obtained were for the isolated disc described by equation (\ref{ecs}). This produces a series of normal modes where the eccentricity propagation is affected only by pressure and is therefore expected to be retrograde. A WKB analysis of the expression was also carried out to act as a check on the numerical method. In fact the eigenvalue of equation (\ref{ecs}) is only dependent on $h$ and the scaled eccentricity distributions are the same for all values of $h$, as a transformation to a new scaled eigenvalue $\omega\to\omega'h^{-2}$ completely eliminates $h$ from the equations.

\begin{figure}
\centering
\includegraphics[angle=0, width=1\columnwidth]{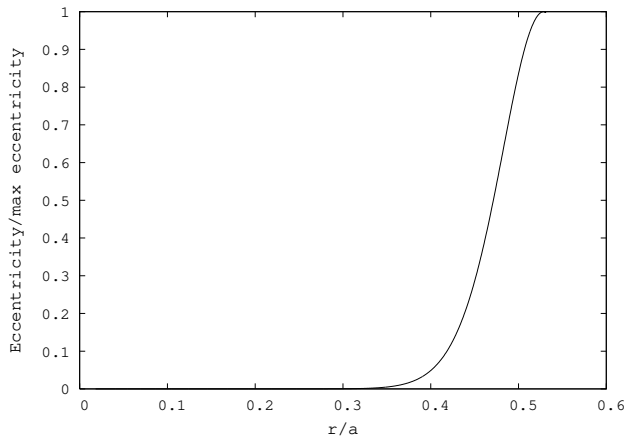}
\caption{The lowest order mode of a disc without the resonance, for OY Carinae with $h=0.003$ and $\alpha_{\rmb}=0.05$.}
\label{fig:mode1}
\end{figure}

The next solution obtained included the effect of the companion as described by equation (\ref{ecb}), using the orbital parameters of the well-studied eclipsing dwarf nova OY Carinae (Wood et al 1989) which has a mass ratio $q=0.102$. The companion is expected to produce a prograde precession of the disc. We find that the gravitational perturbation dominates for the first two modes and overall precession is prograde, but for the remainder of the modes, the precession is retrograde and dominated by pressure. The reason for this is that, as the mode number increases, the eccentricity gradient becomes larger due to the increasing number of nodes and therefore the contribution of the eccentricity gradient term increases. This also affects the damping of eccentricity since this is related to the eccentricity gradient in the same way, and we therefore expect the lowest order mode to be most significant as it is the fastest growing. This lowest order mode has the form shown in Fig. \ref{fig:mode1}, with the greatest eccentricity at the outer edge of the disc.

\begin{figure}
\centering
\includegraphics[angle=0, width=1\columnwidth]{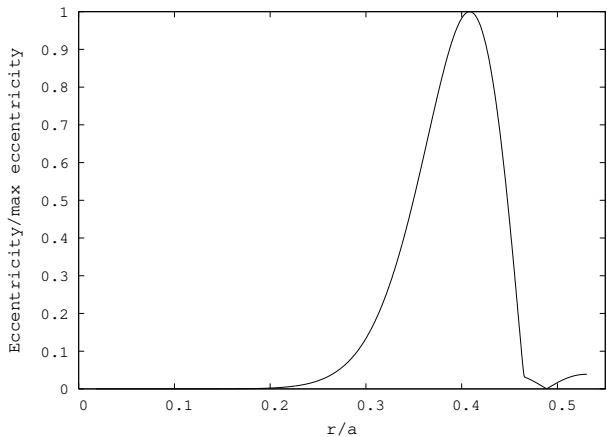}
\caption{The lowest order mode of the full eccentricity equation for OY Carinae with $h$=0.003 and $\alpha_{\rmb}=0.05$.}
\label{fig:moder}
\end{figure}

Finally, we studied the full equation with terms for generation and damping of eccentricity. In the same way as the previous section, the strength of the resonance can be described by a dimensionless parameter which we now write as
\begin{equation}
W=\frac{2\xi\rho\Omega(r_{\mathrm{out}}-r_{\mathrm{in}})}{p}\Bigg|_{\mathrm{res}}.
\end{equation}
For $0.05<q<0.3$, the typical mass ratio range for SU UMa systems (Patterson et al (2005) propose $q<0.35$) then we find the approximate range $20<W<650$ with the physically realistic value $h=0.01$.  The parameter $W$ is dependent on $h$, with $W\propto h^{-2}$, and for the eccentricity distribution shown in Fig. \ref{fig:moder} with $h=0.003$ the eccentricity suppression at the resonant radius is correspondingly larger. The calculated range of $W$ suggests that, in real systems, we are in the region where the local suppression effect of the resonance is important. The addition of the resonance to a real disc indeed leads to a cusp in the eccentricity distribution at the resonant point and the eccentricity distribution, shown in Fig. \ref{fig:moder}, now peaks inside the resonance at about $0.37a$. We find the precession rate is most strongly affected by the mass ratio and semi-thickness, with the bulk viscosity having a negligible effect.

\begin{figure}
\centering
\includegraphics[angle=0, width=1\columnwidth]{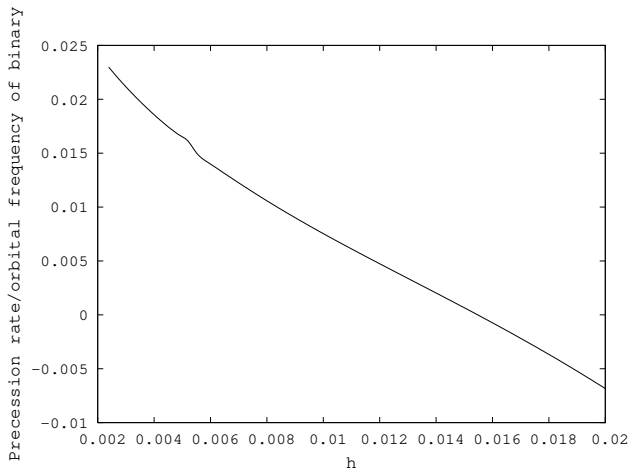}
\caption{The dependence of precession rate on the semi-thickness parameter $h$ for OY Carinae ($q=0.1$).}
\label{fig:omcs}
\end{figure}

Fig. \ref{fig:omcs} shows the effect of varying $h$ in the disc, again using the OY Carinae parameters. The value of the semi-thickness parameter affects the relative importance of the gravitational and pressure terms in the expression; a higher semi-thickness in a hotter disc increases the effect of pressure. This dependence is linear, and for $h>0.015$ the pressure contribution is large enough to make the disc precess retrogradely. OY Carinae has an actual precession rate of $0.0198\Omega_{\rmb}$, and therefore a low value of $H/r$ seems to be more accurate, with a best-fit value $h\simeq0.0035$. This is significantly lower than the value expected from the definition of $h$ in equation (\ref{heq}) which, for reasonable values of primary mass and accretion rate, is about 0.01.

\begin{figure}
\centering
\includegraphics[angle=0, width=1\columnwidth]{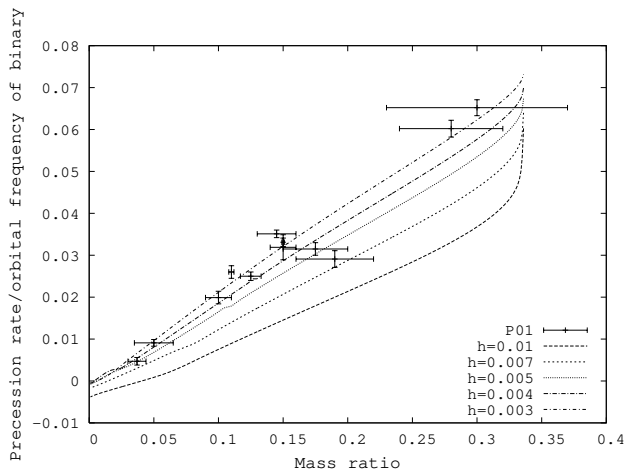}
\caption{Precession rate as a function of mass ratio for different disc semi-thickness parameters. P01 labels data points from Patterson (2001).}
\label{fig:preq}
\end{figure}

We then calculated the variation of precession rate with $q$ for several values of $h$, shown in Fig. \ref{fig:preq} with the observational data from Patterson (2001). Once again the lower values of $h$ produce more accurate agreement, with $h\simeq0.003$ the most consistent value, although the observational errors are sometimes significant. We have assumed that the effect of changes in the ratio $r_{\mathrm{in}}/a$ are unimportant.

Fig. \ref{fig:preq} also shows a linear relation over most of the curve, with deviations at high and low $q$. That at high $q$ is due to the effect of a resonance very close to the disc edge, and is strongly affected by the implementation of the boundary conditions in the pressure and surface density distributions. Over most of the range, and particularly where most observed superhump systems lie, the relation is linear. A best-fit to the linear part of the curve for $h=0.003$, the closest value to the observations, has the relation $\epsilon=(0.2076\pm0.0003)q-(4.1\pm0.6)\times10^{-4}$. This differs from the Patterson (2001) relation by having a small offset from the origin, which is due to pressure-induced retrograde precession of a stable eccentric mode in systems with very low $q$. An isolated system has retrograde eccentric modes, but in the absence of resonant excitation the viscosity damps these away and no long-term eccentricity develops. Otherwise the two relations agree reasonably well.

\subsection{Solutions of the full eccentricity equation-growth}

\begin{figure}
\centering
\includegraphics[angle=0, width=1\columnwidth]{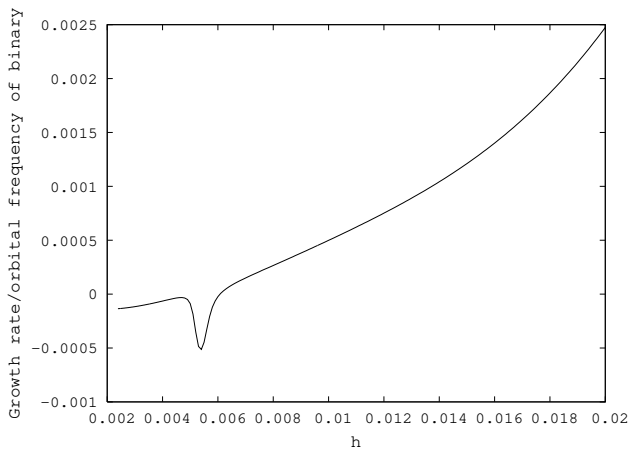}
\caption{Growth rate as a function of semi-thickness for OY Carinae with $\alpha_{\rmb}=0.1$.}
\label{fig:hgro}
\end{figure}

\begin{figure}
\centering
\includegraphics[angle=0, width=1\columnwidth]{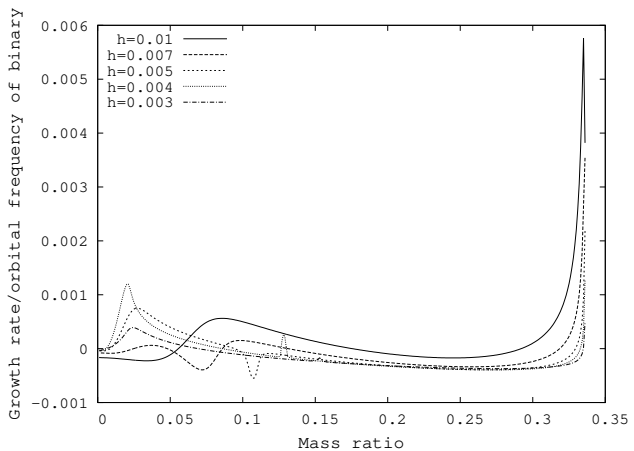}
\caption{Growth rate as a function of mass ratio for different semi-thickness parameters. Bulk viscosity $\alpha_{\rmb}=0.1$.}
\label{fig:qgro}
\end{figure}

Having established the precession rates for the eccentric discs, we then looked at the growth rates of the eccentric modes in the disc. We find that this is most strongly affected by the bulk viscosity and mass ratio, although the semi-thickness also has some effect. Fig. \ref{fig:hgro} shows the dependence of the growth rate on the semi-thickness $h$ for the OY Carinae system with a bulk viscosity $\alpha_{\rmb}=0.1$. Here the eccentric mode is unstable at large $h$, but starts to become stable at the low values of $h$ consistent with the precession rates discussed above. This behaviour is not general, however, as shown by Fig. \ref{fig:qgro} which shows the growth rate as a function of $q$ for different mass ratios. The sharp peaks and troughs are due to changes in the eccentricity at the resonance depending on the eccentricity distribution at the outer edge of the disc: at certain values of the parameters a peak can become `trapped' in the outer part of the disc and lead to a change in the eccentricity at the resonance. For $\alpha_{\rmb}=0.1$, the general behaviour of growth rate shows stability over much of the parameter range for all values of $h$.

\begin{figure}
\centering
\includegraphics[angle=0, width=1\columnwidth]{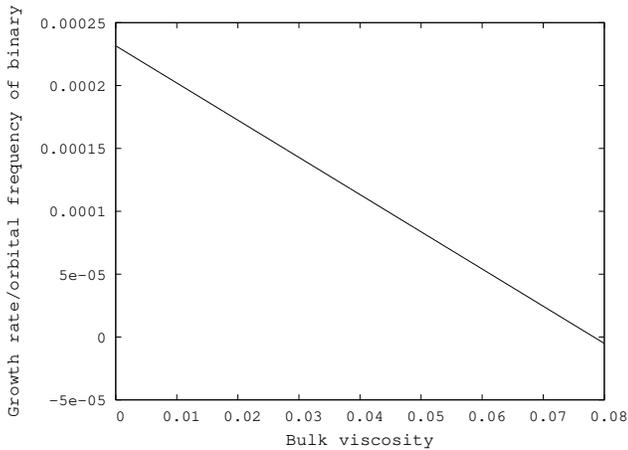}
\caption{Growth rate as a function of bulk viscosity for OY Carinae with \textit{h}=0.003}
\label{fig:agro}
\end{figure}

To determine where the modes are unstable, we therefore consider the dependence of the growth rate on the bulk viscosity, shown in Fig. \ref{fig:agro}. This, as expected from the growth rate equation (\ref{amr}), is approximately linear and gives an unstable mode for $\alpha_{\rmb}<0.079$. To obtain a reasonable value of the growth, however, we need a much lower value of the bulk viscosity, closer to $\alpha_{\rmb}\simeq0.01$. Superhumps are typically observed within a day of the start of a superoutburst, and reach their maximum amplitude within two days (Kato 1996). In terms of a typical orbital period of about 2 hours, we have a characteristic growth time $(-\pi\Im(\omega))^{-1}$ for an increase by a factor e which, for the highest growth rate in Fig. \ref{fig:agro} (without any damping), leads to growth by only a factor 1.02 over a day. This extremely low growth rate is clearly due to the strong suppression of the eccentricity at the resonance.

\section{Summary and discussion}
\label{diss}

We have shown that the effect of including the 3:1 Lindblad resonance in the dynamics of a two-dimensional eccentric disc leads to a suppression of the eccentricity distribution at the resonant point, which has a very important effect on the behaviour of the disc. The precession relation obtained, given by equation (\ref{amd}) and replacing the dynamical equation (\ref{pra}), can produce a much more accurate fit to the observational data than the simple particle precession rate at the 3:1 resonance. Numerical evaluation of the integrals in equation (\ref{amd}) for typical eccentricity distributions shows that the retrograde pressure terms are only $\sim1$ per cent of the dynamical term, and so the dominant effect in reducing the precession rate from the particle precession is the averaging over the disc rather than the retrograde precession. The obtained eccentricity distributions peak at about $r\simeq0.37a$, the value which Patterson (2001) suggested to match the observational data with a dynamical precession. This appears to be an explanation for the observed disc precession rates being so much lower than the expected dynamical precession at the resonance. 

However, to obtain an accurate fit to the data, we have to consider a value of the semi-thickness parameter $h$ which is much lower than the value we would expect. It is difficult to reconcile $h=0.003$ with the value $h\simeq0.01$ which we expect from the standard $\alpha$-disc model, as this requires the disc to be substantially thinner and cooler. This also leads to the associated problem with the growth rates, since a lower $h$ increases the parameter $W$ and so decreases the growth rate, as shown in Fig. \ref{fig:qgro}.

It is possible that these discrepancies between expected parameters and those required to reproduce the observations are due to the simplifications involved in the model. Three-dimensional effects may be significant, as would be the introduction of a shear viscosity. The general behaviour of eccentricity will become non-linear once the linear instability described by Lubow (1991) has started to take hold, and this will modify the eccentricity distribution away from the linear results obtained here. In particular, non-linear effects will be most significant where $\partial E/\partial r$ is large, and therefore we would expect them to be important in the region of the cusp. This might have the effect of smoothing out the cusp and therefore reducing the suppression of eccentricity at the resonance, but further non-linear studies are required to investigate this.

An estimation of the additional effects can be made by comparing the expressions obtained above with the non-linear eccentricity equation obtained by Ogilvie (2001). Starting from the three-dimensional, non-linear eccentricity equation obtained there and working in the linear limit, we use the values for the numerical constants obtained by numerical solution of the associated structure equations in that paper and obtain the following eccentricity equation to compare with the complex version of equation (\ref{ecs}) for a steady $\alpha$ disc:
\begin{gather}
\label{enl}
2r\Omega\frac{\partial E}{\partial t} = \frac{0.06+1.262\rmi}{r^{2}\Sigma}\frac{\partial}{\partial r} \left(  P r^{3} \frac{\partial E}{\partial r} \right)+\nonumber \\(-0.306+0.238\rmi)P \frac{\partial E}{\partial r}+(-0.531+0.938\rmi)\frac{E}{\Sigma} \frac{P}{r}.
\end{gather}
Equation (\ref{ecs}) takes the following form once the power-law form for the pressure has been used to calculate the derivative:
\begin{equation}
2r\Omega\frac{\partial E}{\partial t} = \frac{1}{r^{2}\Sigma}\frac{\partial}{\partial r} \left( (\alpha_{\rmb}+\rmi\gamma) P r^{3} \frac{\partial E}{\partial r} \right)+(-1.5\rmi)\frac{E}{\Sigma} \frac{P}{r}.
\end{equation}
The effective bulk viscosity is slightly reduced from the shear viscosity used (0.1), a consequence of the viscous overstability discovered by Ogilvie (2001) which may help to interpret the low values of $\alpha_{\rmb}$ required for an unstable mode. Introducing a shear viscosity to the problem would allow the overstability to take effect and might increase the growth rates to more realistic levels. The value of 1.262, which is the effective adiabatic index, is slightly reduced but not significantly so, which would tend to reduce the effects of pressure. Since we do not believe these to be very significant in determining the precession of the lowest order mode, it is likely that any effect of this change would be due to changes in the shape of the mode itself. 

The most significant difference between the two expressions is the term proportional to $E$ alone, which has changed sign with the imaginary part of the non-linear expression now positive rather than negative and now describes a prograde pressure-driven precession. This term may have the effect of increasing the value of $h$ which fits the observational data, since it will tend to increase the precession rate of the disc. Further investigation is certainly required to investigate this effect and determine whether a more realistic value of $h$ can be obtained. There is also a completely new term proportional to $\partial E/\partial r$, which may have an entirely new effect on the disc dynamics.

Other refinements to the model may lead to further changes in the precession rates. Numerical solutions of our equations with a varying outer disc radius confirm that these variations, which are expected to occur during superoutbursts, do not have a significant impact on the precession rate. The form of the eccentric mode in Fig. \ref{fig:moder} suggests that the reason for this is the fact that the precession rate in equation (\ref{amd}) is dominated by the part of the mode inside the 3:1 resonance. Variations in the surface density and pressure profiles may be more significant, and observations and simulations suggest that there is a build-up of mass at the outer edge of the disc during superoutbursts (for example see Horne \& Marsh 1986). It is difficult to estimate the precise effect of these variations, however, since they may effect the eccentricity distribution in ways which can only be determined by full solution of the eccentricity equation.

\section{conclusions}
\label{conc}

\begin{enumerate}
\item It is possible to describe the dynamics of an eccentric 2D accretion disc using a single equation which contains information about the eccentricity distribution of the disc, its precession and propagation, and the resonant growth and viscous damping of the eccentricity.
\item By including the 3:1 Lindblad resonance as a delta-function in the eccentricity equation, we find that the resonance suppresses eccentricity at the resonant point itself and leads to a deep cusp in the eccentricity distribution of the fastest-growing mode. This in turn leads to a certain amount of suppression of the eccentricity growth.
\item By solving the eccentricity equation, it is possible to obtain a more accurate fit to the observations of superhumps than has previously been managed using dynamical estimates alone. The closest fit, however, is obtained for values of the characteristic disc semi-thickness of about $h=0.003$, much lower than the value $h=0.01$ which we expect.
\item Further non-linear and three-dimensional calculations are required to obtain a more complete and accurate picture of the dynamics of eccentric discs in SU UMa systems.
\end{enumerate}

\section*{acknowledgements}
The authors would like to thank Steve Lubow for helpful discussions, and the referee, James Murray, for helpful comments which improved the paper. SGG acknowledges the UK PPARC for a research studentship.

\end{document}